\documentstyle[multicol,prb,aps,epsf]{revtex}

\begin{document}

\draft

\title{Diffusion Monte Carlo study of circular quantum dots}

\author{Francesco Pederiva}
\address{Dipartimento di Fisica and INFM, Universit\`a di Trento, I-38050 Povo,Trento, Italy}
\author{C. J. Umrigar}
\address{Cornell Theory Center, Cornell University, Ithaca, NY 14853}
\author{E. Lipparini}
\address{Dipartimento di Fisica and INFM, Universit\`a di Trento, I-38050 Povo,Trento, Italy}
\date{\today}
\maketitle

\begin{abstract} We present ground and excited state energies obtained from
Diffusion Monte Carlo (DMC) calculations, using
accurate multiconfiguration wave functions,
for $N$ electrons ($N\le13$) confined to a circular quantum dot.
We compare the density and correlation energies to the
predictions of local spin density approximation theory (LSDA), and Hartree-Fock
theory (HF), and analyze the electron-electron pair-correlation functions. The DMC estimated
change in electrochemical potential as a function of the number of electrons in
the dot is compared to that from LSDA and HF calculations.
Hund's first rule is found to be satisfied for all dots except $N=4$ for which
there is a near degeneracy.
\end{abstract}


 \begin{multicols}{2}

\section{Introduction}
Modern microfabrication technology is capable of making quantum dots\cite{Kas92,Ash96} that
are sufficiently small that they contain only a small number of mobile electrons.
There has been much interest in studying the
atomic-like properties of these dots
with tunnel conductance\cite{Mei90}  and capacitance\cite{Ash92} experiments.
The ground states of clean circular dots
exhibit shell structure and believed to obey Hund's first rule\cite{Tar96,Aus99}.
The shell structure is particularly evident in
measurements of the change in electrochemical potential due to the addition of
one extra electron
$\Delta_N=\mu(N+1)-\mu(N)$ where $N$ is the number of electrons in the dot, and
$\mu(N)=E(N)-E(N-1)$ is the electrochemical potential of the system.
Theoretical predictions of $\Delta_N$ and the excitation energy spectrum require
accurate calculations ground-state and excited-state energies.
Exact diagonalization studies\cite{Eto97,Maksym90} are accurate for a very small number
of electrons, but the number of basis functions needed to obtain a given
accuracy and the computational cost grow very rapidly with electron number.
In practice they have been used for up to 8 electrons\cite{Eto97,Maksym90}, but the accuracy is
very limited for all except $N\le 3$.
Hartree\cite{Kum90}, restricted Hartree-Fock (HF), spin-and/or-space unrestricted
Hartree-Fock\cite{Fuj96,Mul96,Yan99} (UHF) and
local spin-density (LSDA) and current density functional methods\cite{Kos97,Hir99,Fer94},
give results that are satisfactory for a qualitative understanding of some
systematic properties. However, comparisons with exact results show
discrepancies in the energies that are substantial
on the scale of energy differences. An advantage of the approximate
approaches is that no serious size and geometry constraints are
imposed.

In this paper we employ the Quantum Monte Carlo (QMC) method (both variational
Monte Carlo (VMC) and diffusion Monte Carlo (DMC)) because they yield
very accurate energies at a computational cost that grows relatively modestly
with the number of electrons.
The statistical error of these calculations can be made small, even for dots
with several tens of electrons, within a reasonable amount of
computer time on a modern workstation.  In addition to the statistical error
there is a systematic
error due to using the fixed-node approximation.  This error can be reduced
by optimizing the trial wavefunctions.  For the trial wavefunctions used in
the present work, the fixed-node errors are small compared to the errors
of other approximate methods.  This is demonstrated by performing internal
checks within the method and by comparing to the few energies, available from
exact diagonalization studies\cite{Eto97,Maksym90} for small dots, that are accurate
enough to make a meaningful comparison.
Hence our results can be regarded as a benchmark to assess the accuracy of
other approximate methods.  In particular we find that, in contrast to the
situation with atoms, the energies
obtained from the LSDA method are considerably more accurate than those from
the HF method.  The same is true for the spin-densities in those cases where
the LSDA wavefunctions are eigenstates of the total spin operator $\hat S^2$.

Earlier QMC calculations on quantum dots include VMC calculations
for circular dots~\cite{Har99} and DMC calculations for 3-dimensional dots~\cite{Shu99}.
Fixed-phase DMC has been applied to dots~\cite{Bol96} with $N \le 4$.
Path integral Monte Carlo calculations have been performed~\cite{Egg98} for dots
with $N \le 8$ but the results of these calculations bear no resemblance to
either our results or those from exact diagonalization~\cite{Eto97,Maksym90}.

\section{Computational Method}
\subsection{Hamiltonian}
\label{hamiltonian}
The usual model~\cite{Ash96} for a disk-shaped vertical quantum dot is
a 2-dimensional system of $N$ electrons moving in the $z=0$ plane,
confined by a parabolic lateral confining potential $V_{\rm con}({\bf r})$.
The  Hamiltonian is
\begin{equation}
H=\sum_{i=1}^N(-{\hbar^2\over2 m^*}\nabla_i^2+ V_{\rm con}({\bf r}_i)) +
{e^2\over\epsilon}\sum_{i<j}^N{1\over \vert{\bf r}_i-{\bf r}_j\vert}~~.
\label{eq1}
\end{equation}
In Eq. (\ref{eq1}), $m^*$ is the electron effective mass,
and $\epsilon$ is the dielectric constant of the semiconductor.
In  the following (if not explicitly specified otherwise) we
will use effective atomic units, defined by $\hbar=e^2/\epsilon=m^*=1$. In this
system of units, the length unit is the Bohr radius $a_0$ times $\epsilon m_e/m^*$,
and the energy unit is the Hartree times $m^*/(m_e \epsilon^2)$.
For the GaAs dots
we consider here, $\epsilon=12.4$ and $m^*=0.067 m_e$, and the effective
Bohr radius $a_0^*$ and effective Hartree H$^*$ are
$\simeq 97.93 \AA$ and $\simeq 11.86$ meV
respectively.
In this first application of the method, we will consider
circular dots with $N\le 13$, and a parabolic potential, $V_{\rm con}({\bf r})=m^* \omega^2r^2/2$
($\hbar\omega$=0.28 H$^* = 3.32$ meV),
which should approximate the
experimental situation in Ref.~\onlinecite{Tar96}.
Extensions of the calculation to $N > 13$, magnetic field $B\ne0$ and
a non-parabolic confining potential are in progress.

Comparison of energies and other quantities with those in the literature
are complicated by the fact that various authors use different values for
the parameters, $m^*, \epsilon, \omega$ in the Hamiltonian.
Note however, that two Hamiltonians, $H_1$ and $H_2$, characterized by
$m^*_1, \omega_1$ and $\epsilon_1$ and $m^*_2, \omega_2$ and $\epsilon_2$, respectively,
must have the same energy spectrum aside from a multiplicative scale factor, i.e.,
$E_{1i}/E_{2i} = m^*_2/m^*_1 = \omega_1/\omega_2 = \epsilon_2/\epsilon_1$, where $i$
labels the energy states of a given Hamiltonian.
An interesting aspect of quantum dots is that it is possible to tune, $\lambda$,
the dimensionless ratio of the Coulomb interaction strength to
the confining potential $\lambda=[e^2/(\epsilon l_0)]/\hbar \omega$,
where $l_0 = \sqrt{\hbar/(m^*\omega)}$, thereby allowing one to study both
weakly interacting and strongly interacting cases.
Our present calculations are for $\lambda=1.89$.

\subsection{Quantum Monte Carlo Methods}
One advantage of the QMC methods is that no restriction is placed on the form
of the trial wavefunction.  In the VMC method, Monte Carlo integration is used
to calculate the many-dimensional integrals, and the parameters in the trial
wavefunction can be varied to minimize the energy or the fluctuations of the
local energy.  More accurate results can be obtained from the fixed-node
diffusion Monte Carlo method which
projects, from an
antisymmetric trial wave function, the
lowest energy state, consistent with the boundary condition of preserving the
same nodal surface.
In the limit that the trial wavefunction has the correct nodes, fixed--node
DMC yields the exact energy with only a statistical error
that can be made arbitrarily small by increasing the number of Monte Carlo steps.
A detailed description of our implementation can be found in Ref.~\onlinecite{Umr93}.
The fixed-node error is usually small compared to the errors from other methods
but it is unknown except in those cases where exact results are available.

\subsection{Trial Wavefunctions}

The errors of VMC and fixed-node DMC calculations depend on the
quality of the trial wavefunctions.
The trial wavefunctions we use have the form
\begin{equation}
\Psi({\bf R})_{L,S}=
\exp[\phi({\bf R})]\sum_{i=1}^{N_{\rm conf}} \alpha_i\Xi_i^{L,S}({\bf R})~~,
\label{eq2}
\end{equation}
where ${\bf R}=\{{\bf r}_1\cdots{\bf r}_N\}$ are the coordinates of the $N$
electrons in the dot, and the $\alpha_i$ are variational parameters.
The configuration-state-functions, $\Xi^{L,S}$, are eigenstates of the total angular
momentum $\hat{L} \equiv \hat{L}_z$ with eigenvalue $L$ and of the total spin $\hat{S}^2$ with eigenvalue
$S(S+1)$, and have the following form:
\begin{equation}
\Xi^{L,S}_i=\sum_{j=1}^{m_i} \beta_j D^\uparrow_j D^\downarrow_j~~,
\label{eq3}
\end{equation}
where the $D^\chi_j$ are Slater determinants of spin-up and spin-down electrons,
using orbitals from a local density approximation (LDA) calculation with the same confining potential
and the same number of electrons. The  $m_i$ are the number of determinants in the
$i$--th configuration.
In general the $D^\uparrow_j D^\downarrow_j$ are not eigenstates of $\hat S^2$.
The coefficients $\beta$ in the linear combination of Eq. (3) are fixed by diagonalizing
$\hat S^2$ in that determinantal basis.
For $N \le 13$, the number of configurations, $N_{\rm conf}$ and Slater determinants
$N_{\rm det} = \sum_{i=1}^{N_{\rm conf}} m_i$, appearing in Eqs. 2 and 3
are shown in Table~\ref{tab.excited} and were determined by limiting the
basis space to spin-up and spin-down orbitals with
$|n,l\rangle = |0,0\rangle$ for $N\le2$ dots,
$|n,l\rangle = |0,0\rangle$ and $|0,\pm1\rangle$, for $3 \le N \le 6$ dots,
$|n,l\rangle = |0,0\rangle$, $|0,\pm1\rangle$, $|0,\pm2\rangle$, and $|1,0\rangle$, for $7 \le N\le12$ dots, and.
$|n,l\rangle = |0,0\rangle$, $|0,\pm1\rangle$, $|0,\pm2\rangle$, $|1,0\rangle$, $|0,\pm3\rangle$,
and $|1,\pm1\rangle$ for the $N=13$ dot.
The noninteracting single-particle energy levels are
$\epsilon_{n,l} = (2n+|l|+1)\omega$\cite{FockDarwin}.
Basis states are then built by considering all possible occupations of
open-shell levels.  For example, in the case of the $N=9$ dot,
the first 6 electrons fill the $|0,0\rangle$ and $|0,\pm1\rangle$
orbitals and are considered to be core electrons in a closed shell.
Then, the wave function for the state
$|L=0,S=1/2\rangle$ of the $N=9$ dot has 3 open-shell
electrons, and includes two ($N_{\rm conf}=2$) configuration state functions
which are linear combinations of $m_1=2$ and $m_2=3$ Slater determinants
respectively.

The function $\exp[\phi]$ in Eq. (\ref{eq2}) is a generalized Jastrow
factor of the form used in Ref.~\onlinecite{Bal92},
\begin{eqnarray}
\begin{array}{l}
\phi(R)=\displaystyle\sum_{i=1}^{N}\left[\sum_{k=1}^{6}\gamma_k J_0\left(
\frac{k\pi r_i}{R_c}\right)\right] +\\
\displaystyle\sum_{i<j}^N\frac{1}{2}\left(\frac{a_{ij} r_{ij}}{1+b(r_i)r_{ij}}+
\frac{a_{ij} r_{ij}}{1+b(r_j)r_{ij}}\right)~,
\end{array}
\end{eqnarray}
where
\begin{equation}
b(r)=b_0^{ij}+b_1^{ij}\tan^{-1}[(r-R_c)^2/2R_c\Delta].
\end{equation}
It explicitly includes one- and two-body correlations and effective
multi-body correlations through the spatial dependence of $b(r)$.
The quantity $R_c$ represents an ``effective'' radius of the dot,
and has been assumed to be equal to $1.93\sqrt{N}$.
The $b_0$ and $b_1$ parameters depend only on the {\it relative} spin
configuration of the pair $ij$.
The parameters $a_{ij}$ are fixed in order to satisfy the
cusp conditions, that is, the condition of finiteness of
the local energy $\hat H\Psi/\Psi$ for $r_{ij}\rightarrow 0$. For a two
dimensional system, $a_{ij}=1$ if the electron pair $ij$ has antiparallel
spin, and $a_{ij}=1/3$ otherwise.
The dependence of $a_{ij}$ on the relative spin orientation of the electron
pair introduces spin-contamination into the wavefunction.
However, the magnitude of the spin contamination and its effect on the
energy has been shown to be totally negligible in the case of well optimized
atomic wavefunctions~\cite{Chi98} and we expect that to be true here as well.

The coefficients $\gamma_k$ in the one--body term,
the coefficients $\Delta$, $b_0$, and $b_1$ in the two--body term and
the coefficients $\alpha_i$ multiplying the configuration state functions
are optimized by minimizing the variance of the local energy~\cite{UWW88}.
The resulting wavefunctions had rms fluctuations of the local energy that
range from 0.021 H$^*$ for $N$=2 to 0.255 H$^*$ for $N$=13.

\section{Results}
Using these optimized wave functions for importance sampling, we perform
fixed--node diffusion Monte Carlo calculations.
We attempt to establish the accuracy of the fixed--node energies obtained with
our trial wavefunctions
by comparing them to energies from exact diagonalization studies and also
by performing internal checks within our calculations.
Unfortunately, although there exist several papers on exact diagonalization~\cite{Eto97,Maksym90},
the results are usually presented in plots, rather than in tables.
The only number we know of is in Ref.~\onlinecite{Har99}, who give an energy
of 26.82 meV for $N=3$, that they credit to Hawrylak and Pfannkuche\cite{Maksym90}.
Starting from a single Slater determinant of LDA orbitals (constructed from the
spin-up and spin-down $|n,l\rangle = |0,0\rangle$ states and the
spin-up $|0,1\rangle$ spin up) state,
we obtain a fixed-node DMC energy of 26.8214(36) meV, using
their model parameters ($m^*=0.067 m_e, \epsilon=12.4$ and $\hbar\omega=3.37$ eV),
which is indistinguishable from the exact energy to the number of digits quoted.
We attempted also to estimate the fixed-node error by varying the orbitals in the
determinants and by varying the number of determinants.
For $N=7$, QMC calculations using local density approximation (LDA) and
local spin-density approximation (LSDA) orbitals were performed.
The LSDA orbitals yielded better VMC results (the energy was lowered by
57 mH$^*$, and the fluctuations of the local energy by 7 mH$^*$)
but the DMC energies were unchanged within statistical uncertainty.
We checked the dependence of the energy on the number of configuration state functions
for the first excited state of the $N=9$ dot.
Somewhat to our surprise, energies of the one-configuration (three-determinant)
and the two-configuration (five-determinant) wavefunctions agreed to within
1 mH$^*$, not only within DMC but also within VMC.

\subsection{Ground-state energies}
The ground state energies are listed in Table~\ref{tab.ground} and compared with
results of HF, and LSDA calculations using the Tanatar-Ceperley parametrization for
the correlation energy~\cite{Tan89}.  The HF energies are 0.12 to 0.97 H$^*$ higher
than the DMC energies whereas the LSDA energies are only 0.021 to 0.042 mH$^*$
(0.25 to 0.50 meV) higher.
In contrast, in atoms and molecules the Hartree-Fock total energy is considerably better
than the LSDA total energy.
There are two likely reasons for this difference.  First, Hartree-Fock treats exchange exactly
while completely ignoring correlation, whereas in LSDA both exchange and correlation
are approximated.  In atoms and molecules, the exchange energy, $E_x$, is much larger than
the correlation energy, $E_c$, but for the dots it is not, e.g. $E_x/E_c \approx 30$ for
a Neon atom but $E_x/E_c \approx 4$ for a $N=10$ dot.  The second reason is that
the dots are more homogeneous than atoms or molecules and so the local-density approximations
to $E_x$ and $E_c$ work better.
Note also that the HF errors increase monotonically with electron number but
the LSDA errors do not show any obvious trend.

\subsection{Excited-state energies}
In Table~\ref{tab.excited} we list the low-lying excitation
energies for the $N=4\dots11$ dots.
Koskinen et al. (Ref.~\onlinecite{Kos97}) find that the lowest
excitation energies from LSDA calculations are 11.5 mH$^*$ and 2.31 mH$^*$
for $N=8,10$ dots respectively, whereas our DMC calculations show that the
corresponding lowest excitation energies are 22 mH$^*$ and 2 mH$^*$.
They claim that these lowest excited states have a spin density wave
even though $S=0$, but
in fact this is just an artifact due to these LSDA wavefunctions not being
eigenstates of $\hat S^2$, as pointed out by Hirose and Wingreen~\cite{Hir99}.
For both the $N=8$ and the $N=10$ dots, the first excited state LSDA wavefunctions
are in fact linear combinations of $(L,S)=(2,1)$ and the  $(L,S)=(2,0)$
wavefunctions.
In general, the single-determinant LSDA wavefunctions are eigenstates of $\hat{S}_z$,
but they
are eigenstates of $\hat{S}^2$ only when
$\vert\hat{S}_z\vert$ has the maximum value consistent with filling the lowest $N/2$
orbitals and the exclusion principle.
In other cases it is necessary to
have more than one determinant in order to have the correct spin symmetry.
Single-particle levels in a parabolic potential with the same value of $2n+\vert l
\vert+1$
are degenerate.  However, the self-consistent LSDA potential is not parabolic
and consequently of two levels with the same $2n+\vert l \vert+1$, the LSDA orbital
with the larger value of $\vert l \vert$ is lower than the other.
This serves to explain the ordering of levels for the dots where $N$ differs from
a closed shell by one.  For example, the $(L,S)=(2,1/2)$ state lies lower than
the $(L,S)=(0,1/2)$ state in the $N=7$ dot but the order is reversed in the $n=11$
dot because the $(n,l)=(0,\pm 2)$, LSDA single-particle level lies below the
$(n,l)=(1,1)$
level.

\subsection{Change in electrochemical potential}
The DMC estimates for the change in electrochemical
potential $\Delta_N$  (in meV) as a function of $N$
are reported in Fig.~\ref{fig.eadd} together with those
from LSDA and HF calculations.
We see structures and peaks at electron numbers 2,4,6,9 and 12 in
agreement with the experiments of Ref.~\onlinecite{Tar96}. In the independent-particle model with a
parabolic potential, $\Delta_N$ has peaks of magnitude $\omega$
at $N=2,6,12...$, corresponding of closed shells, and is 0 elsewhere.
Additional features are due to the electron--electron interaction.
It is difficult to make a more detailed comparison between experiments and theory because
of uncertainties in the Hamiltonian.
In particular, the external potential may not be strictly parabolic and our
assumption that $\omega$ is independent of $N$ may not be an accurate description
of the experimental situation.

\subsection{Correlation energies}
In Fig.~\ref{fig.ecorr} we plot the DMC correlation energy calculated
as the difference between
the DMC energy and the HF energy~\cite{DFT_QC_diff} as a function of the electron number $N$.
The dashed line indicates the LSDA correlation energy.  From the figure one sees that
LSDA overestimates the correlation energy by 10\%-15\% almost independently of $N$.
The LSDA overestimate of the correlation energy is smaller than in atoms and jellium spheres,
where it is as much as 100\% (Ref.~\onlinecite{Per81})
and 30\% (Ref.~\onlinecite{Bal92}), respectively.

\subsection{Hund's first rule}
{}From Table~\ref{tab.excited} we see that Hund's first rule, according to which the total spin of
the ground state takes the maximum value consistent with electrons being in the same shell
and the exclusion principle, is satisfied for all values of $N$ studied in this
work, except for $N=4$.  For $N=4$ the $\vert L,S\rangle = \vert 0,0\rangle$ state is just
2.2 mH$^*$ or 0.026 meV lower than the $\vert 0,1\rangle$ state, so a small change in the Hamiltonian,
e.g., an increase in the spring-constant of the confining potential, $\omega$, could
alter the ordering of these two states.
Our result for the $N=4$ dot is in qualitative agreement with the QMC results of Bolton\cite{Bol96}
but they find that the the singlet state is lower than the triplet by a larger amount (1.5 meV)
than we do, for Hamiltonian parameters that are close to, but not equal to, the the ones we use.
However, our result disagrees with our LSDA calculations which find no violations of Hund's rule for
$N\le13$ as well as the earlier LSDA calculations~\cite{Kos97} which found that, for dots
with even $N$, Hund's rule is satisfied for $N\le22$, but violated for $N=24$.
On the other hand spin-and-space unrestricted Hartree Fock
(sS-UHF) calculations~\cite{Yan99} predict that Hund's rule is violated not only for $N=4$
but also for $N=8$ and $N=9$.  It should be noted that the sS-UHF were performed for
a smaller value, $\lambda=1.48$, 
of the dimensionless ratio of the Coulomb interaction strength to
the confining potential, defined in Section~\ref{hamiltonian},
than our calculations which were for $\lambda=1.89$.
Since, according to Ref.~\onlinecite{Yan99}, Hund's rule violations are less likely for smaller
values of $\lambda$, it is clear that the difference is not due to the different value of $\lambda$.
Experimental evidence indicates that Hund's rule is satisfied for $N=4$ circular dots~\cite{Tar96,Aus99}
but that a small elliptical deformation is sufficient for the singlet and triplet energies to
cross~\cite{Aus99}, thereby confirming our finding that the two states are very close in energy.
Given the uncertainty in the experimental Hamiltonian and the near degeneracy of the two states,
it is not surprising that we find that the
the singlet state is lowest whereas the experimental finding is that the triplet is lowest for the circular dot.
The results of two exact diagonalization studies are also relevant in this context:
Eto~\cite{Eto97} found that a small magnetic field is sufficient to switch the order of
the states whereas Hirose and Wingreen~\cite{Hir99} find that a small quartic term in the Hamiltonian has
the same effect.

\subsection{Spin densities}
In Fig.~\ref{fig.spindens} we compare the spin densities, $\rho_{\uparrow}$ and
$\rho_{\downarrow}$, and the magnetization,
$m(r)=\rho_{\uparrow}(r)-\rho_{\downarrow}(r)$ for the $N=9$ ground state
obtained from DMC and LSDA.
In this case the LSDA wavefunction is an eigenstate of $\hat S^2$.
The agreement of the curves is impressive and extends to the whole region of $r$,
including the edge, where the density gradients are large.
The same kind of agreement with LSDA was also obtained
in the case of variational Monte Carlo densitites\cite{Bal92} of jellium spheres.
In general, it appears that LSDA gives accurate spin densities in those cases that the
Kohn-Sham wavefunction has the correct spin symmetry.
In contrast the HF spin densities show much larger oscillations than the DMC spin densities.
The same behavior has previously been noticed for atoms, but to a much lesser degree~\cite{Fil96}.

Yannouleas and Landman plot in Fig. 2 of Ref.~\onlinecite{Yan99} the charge density of a closed-shell $N=6$
dot obtained from a spin-and-space unrestricted Hartree Fock (sS-UHF) calculation.
They find that the charge density
of a dot with a dimensionless interaction strength of $\lambda=1.48$ has a
non circular charge density that they refer to as a Wigner crystallized state,
although the usual definition of Wigner crystallization refers to the occurrence
of long-range order in the 2-body density rather than short-range order in the
1-body density.
Since the ground state of the $N=6$ dot is of $^1S$ symmetry, it is apparent that
the density must be circularly symmetric and their result is an artifact of their
computational method.  In this context it should be noted that for very large
values of $\lambda$ one cannot immediately rule out the possibility that
the single-particle picture breaks down completely and that the ground
state is not of $^1S$ symmetry.  Also, in the presence of a strong magnetic field
the single-particle levels will reorder and the ground
state need not have $^1S$ symmetry.  Finally, it should be noted that other authors~\cite{Rei98}
have considered models for dots in which the confining potential itself
can deform and therefore not be circularly symmetric.  In this case, of
course, the ground state density of the $N=6$ dot need not be circularly symmetric either.

\subsection{Pair correlation functions}
In Fig.~\ref{fig.paircorr} we show the spherical average of the electron-electron pair correlation functions
$g_{\sigma_1,\sigma_2}({\bf r}_1,{\bf r}_2)$ in the $N=9$ case. The different behavior for
pairs with parallel and antiparallel spin is due to the fact that the wavefunction
vanishes when parallel-spin electrons coalesce but not when antiparallel-spin electrons
coalesce.
For $N=9$, it follows from Hund's rule that there are twice as many up-spin electrons
as down-spin electrons. This is reflected in the shape of
the $g_{\uparrow\uparrow}$ and $g_{\downarrow\downarrow}$ curves.

\section{Conclusions}
In conclusion, we have calculated QMC ground-state energies, excitation energies,
correlation energies, change in electrochemical potential due to adding an electron,
spin densities and pair-correlation functions for circular quantum dots with
$N\le13$ electrons and compared them to the corresponding quantities obtained
from HF and LSDA calculations.  We find that HF energies are in error by
0.12 to 0.97 H$^*$ but LSDA energies by only 0.021 to 0.042 H$^*$.
However, even the LSDA energies are not sufficiently accurate to give
reliable excitation energies or changes in the electrochemical potential.
The LSDA correlation energies differ from the DMC ones
by$\simeq$ 10\%-15\%.
Hund's first rule is found to be satisfied for all dots up to $N \le 13$, 
for the Hamiltonian parameter values employed, except for $N=4$ which has
a near degeneracy.
The LSDA spin densities are in remarkably good agreement with DMC for those
cases where the Kohn-Sham wave function is an eigenstate of $\hat S^2$  but
the HF densities have oscillations that are too large.
Finally, the pair-correlation functions may be of utility in constructing
more accurate energy density functionals than the LSDA for 2-dimensional systems.

\section{Acknowledgements}
We thank Ll. Serra for providing us with the HF code,
and M. Barranco for useful conversations.
This work was partially supported by INFM, MURST
and Sandia National Laboratory.
The calculations were carried out on the IBM-SP2 at Cornell Theory Center,
and on the Cray-T3E at CINECA with an ICP-INFM grant.

\begin{minipage}{3.375in}
\begin{table}
\caption{Ground state energies (in H$^*$) and
low-lying excitation energies (in mH$^*$) for $N\le13$ dots.
Also shown are the quantum numbers of the states and
the number of configuration state functions $N_{\rm conf}$ and the
number of determinants $N_{\rm det}$ used in constructing them.
The numbers in parentheses are the statistical uncertainties in the last digit.
}
\begin{center}
\begin{tabular}{cccccr}
$N$ & $L$ & $S$ & $N_{\rm conf}$ & $N_{\rm det}$ & $E$(H$^*$), $\Delta E$(mH$^*$)\\
 \hline
 2 & 0 & 0 & 1 & 1 & 1.02162(7)\\
 \hline
 3 & 1 & 1/2 & 1 & 1 & 2.2339(3)\\
 \hline
 4 & 0 & 0 & 1 & 2 & 3.7135(4)\\
   & 0 & 1 & 1 & 1 & 2.2(6)\\
   & 2 & 0 & 1 & 1 & 41(1)\\
 \hline
 5 & 1 & 1/2 & 1 & 1 & 5.5336(3)\\
 \hline
 6 & 0 & 0 & 1 & 1 & 7.5996(8)\\
 \hline
 7 & 2 & 1/2 & 1 & 1 & 10.0361(8)\\
   & 0 & 1/2 & 1 & 1 & 24(1)\\
 \hline
 8 & 0 & 1 & 1 & 1 & 12.6903(7)\\
   & 2 & 1 & 1 & 2 & 22(1)\\
   & 2 & 0 & 1 & 2 & 24(1)\\
   & 4 & 0 & 1 & 1 & 32(1)\\
   & 0 & 0 & 2 & 3 & 54(1)\\
 \hline
 9 & 0 & 3/2 & 1 & 1 & 15.5784(7)\\
   & 0 & 1/2 & 2 & 5 & 43(1)\\
   & 2 & 1/2 & 2 & 2 & 52(1)\\
   & 4 & 1/2 & 1 & 1 & 67(1)\\
 \hline
10 & 2 & 1 & 1 & 2 & 18.7244(5)\\
   & 2 & 0 & 1 & 2 &  2(1)\\
   & 0 & 1 & 1 & 1 & 22(1)\\
   & 0 & 0 & 2 & 3 & 26(1)\\
   & 4 & 0 & 1 & 1 & 45(1)\\
 \hline
11 & 0 & 1/2 & 1 & 1 & 22.0750(4)\\
   & 2 & 1/2 & 1 & 1 & 14(1)\\
 \hline
12 & 0 & 0 & 1 & 1 & 25.6548(7)\\
 \hline
13 & 3 & 1/2 & 1 & 1 & 29.4942(7)\\
   & 1 & 1/2 & 1 & 1 & 40(1)\\
\end{tabular}
\end{center}
\label{tab.excited}
\end{table}
\end{minipage}

\begin{minipage}{3.375in}
\begin{table}
\caption[]{Comparison of ground state energies (in H$^*$) for the dots with $2\leq N\leq13$
computed by Hartree--Fock, LSDA, VMC and DMC. Also shown are the LSDA errors
in the energy, $\Delta E_{\rm LSDA} = E_{\rm LSDA}-E_{\rm DMC}$, which
are much smaller than the HF errors $E_{\rm HF}-E_{\rm DMC}$.
The numbers in parentheses are the statistical uncertainties in the last digit.
}
\begin{center}
\begin{tabular}{rrrrrr}
N&$E_{\rm HF}$&$E_{\rm LSDA}$&$E_{\rm VMC}$&$E_{\rm DMC}$&$\Delta E_{\rm LSDA}$\\
\hline
2&1.1420&1.04685&1.02205(7)&1.02162(7)&0.02523(7)\\
3&2.4048&2.2631&2.5022(3)&2.2339(3)&0.0292(3)\\
4&3.9033&3.6864&3.7252(4)&3.7135(5)&0.0276(7)\\
5&5.8700&5.5735&5.5473(5)&5.5336(3)&0.0263(7)\\
6&8.0359&7.6349&7.6214(3)&7.5996(8)&0.0353(8)\\
7&10.5085&10.0718&10.0587(9)&10.0361(8)&0.0357(8)\\
8&13.1887&12.7276&12.7119(7)&12.6903(7)&0.0373(7)\\
9&16.1544&15.6190&15.6039(9)&15.5784(7)&0.0406(7)\\
10&19.4243&18.7636&18.7568(9)&18.7244(5)&0.0392(5)\\
11&22.8733&22.1114&22.1128(9)&22.0750(4)&0.0364(4)\\
12&26.5490&25.6756&25.6792(11)&25.6548(7)&0.0208(7)\\
13&30.4648&29.5363&29.5430(14)&29.4942(7)&0.0421(7)
\end{tabular}
\end{center}
\label{tab.ground}
\end{table}
\end{minipage}


\begin{minipage}{3.375in}
\begin{figure}
\vspace{6mm}
\centerline{\epsfxsize=8.0cm \epsfysize=6.5cm \epsfbox{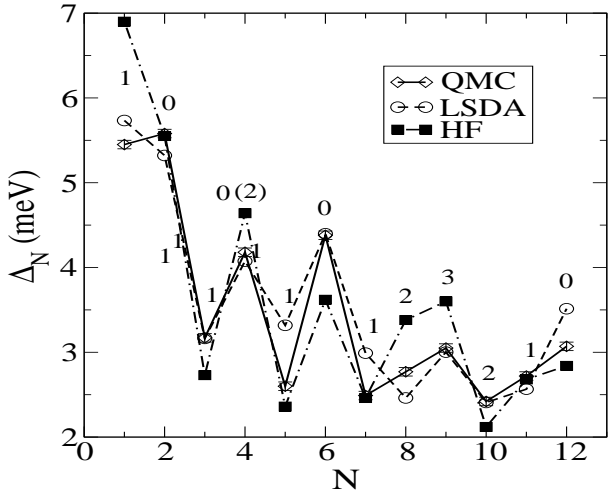}}
\vspace{0.5cm}
\caption{Change in electrochemical potential $\Delta_N$ as a function of the
number of electrons, $N$, in the dot.
The numbers in the plot are the DMC spin polarizations $2S_z$=$N\uparrow-N\downarrow$.
The LSDA and HF spin polarization for the $N=4$ dot is given in parentheses;
for all other $N$ they are the same as for DMC.}
 \vspace{.5cm}
\label{fig.eadd}
\end{figure}
\end{minipage}

\noindent
\vspace{1cm}
\begin{minipage}{3.375in}
\begin{figure}
 \centerline{\epsfxsize=8.0cm \epsfysize=6.5cm \epsfbox{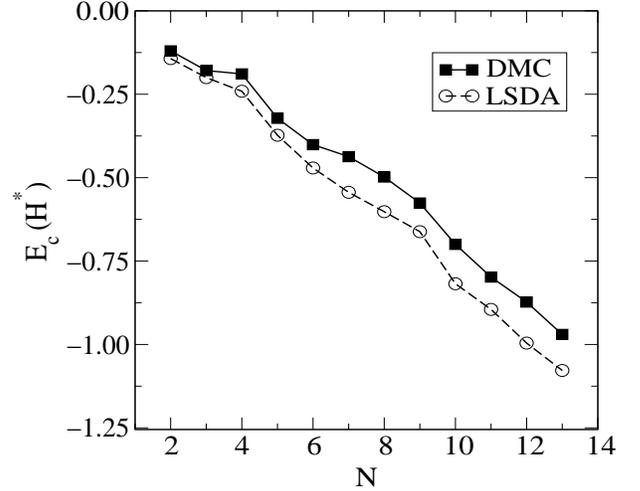}}
 \vspace{0.4cm}
\caption{Correlation energies $E_c$ for circular dots computed with
DMC (filled squares) and LSDA (open circles).}
\vspace{.3cm}
\label{fig.ecorr}
\end{figure}
\end{minipage}

\begin{minipage}{3.375in}
\begin{figure}
 \vspace{9mm}
 \centerline{\epsfxsize=8.0cm \epsfysize=6.5cm \epsfbox{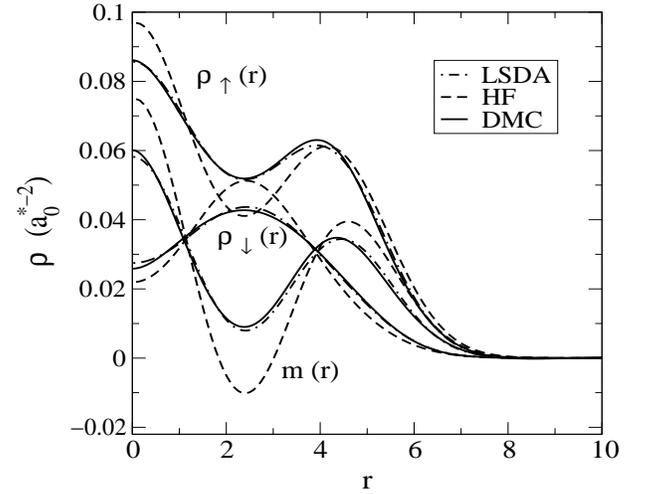}}
 \vspace{0.45cm}
\caption{Spin densities, $\rho_\uparrow(r), \rho_\downarrow(r)$
and magnetization $m(r)=\rho_\uparrow(r)-\rho_\downarrow(r)$
as a function of distance from the center for the ground state
of the $N=9$ dot. Solid lines,
DMC; dotted-dashed lines, LSDA, dashed line, HF.
The LSDA spin densities for this state agree well with the DMC
spin densities but the HF spin densities have considerably
larger oscillations.
}
 \vspace{.3cm}
\label{fig.spindens}
\end{figure}
\end{minipage}

\vspace{1cm}
\begin{minipage}{3.375in}
\begin{figure}
\vspace{6mm}
 \centerline{\epsfxsize=8.0cm \epsfysize=6.5cm \epsfbox{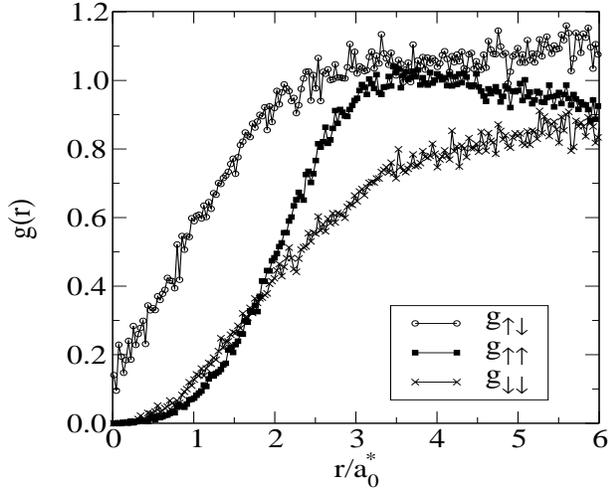}}
 \vspace{0.5cm}
\caption{Electron pair correlation functions,
from variational MC, of the ground state of the $N=9$ dot.
The first electron is at the center of the dot.
Empty circles: $g(r)_{\uparrow\downarrow}$; filled squares:
 $g(r)_{\uparrow\uparrow}$; crosses: $g(r)_{\downarrow\downarrow}$.
}
 \vspace{.3cm}
\label{fig.paircorr}
\end{figure}
\end{minipage}

\end{multicols}


\begin{thebibliography}{99}
\bibitem{DFT_QC_diff} We are neglecting here the small difference between the
density functional and the quantum chemistry definitions of the correlation energy.
\end{thebibliography}
\end{document}